\documentstyle [prl,epsfig,aps,twocolumn] {revtex}
\begin{document}

\newcommand {\lip}   {Li$_{0.9}$Mo$_6$O$_{17}$}
\newcommand {\ef}    {E$_F$}
\newcommand {\kpar}  {{\bf k}$_\parallel$}
\newcommand {\kperp} {k$_\perp$}
\draft

\title{Non-fermi-liquid single particle lineshape of the quasi-one-dimensional
non-CDW metal \lip\ : comparison to the Luttinger liquid}

\author {J.D.\ Denlinger$^1$, G.-H.\ Gweon$^1$, J.W.\ Allen$^1$,  C.G.\
Olson$^2$, J.\ Marcus$^3$, C.\ Schlenker$^3$, and 
L.-S.\ Hsu$^1$\cite{hsu_pa}}

\address {$^1$ Randall Laboratory of Physics, University of Michigan,
Ann Arbor, MI 48109-1120, USA\\$^2$ Ames Laboratory, Iowa State
University, Ames, Iowa 50011, USA\\
$^3$Laboratoire d'Etudes des Propri\'et\'es Electroniques des Solides -
CNRS, BP166, 38042 Grenoble Cedex, France\\}
\date {\today}
\maketitle
\begin {abstract}
We report the detailed non-Fermi liquid (NFL) lineshape of the dispersing
excitation which defines the Fermi surface (FS) for quasi-one-dimensional \lip.
The properties of \lip\ strongly suggest that the NFL behavior has a purely
electronic origin.  Relative to the theoretical Luttinger liquid lineshape, we
identify significant similarities, but also important differences.
\end {abstract}

\pacs {PACS numbers: 71.10.Pm, 71.18.+y, 79.60.-i}

A topic of high current interest and fundamental importance for condensed
matter physics is the possible failure due to electron-electron interactions
\cite{anderson} of the Fermi liquid paradigm for metals.  The paradigm lattice
non-Fermi liquid (NFL) scenario for a metal is the Luttinger liquid (LL)
behavior \cite{haldane} of an interacting one-dimensional electron gas.  The
energy ($\omega$) and momentum ({\bf k}) resolved single particle spectral
function A({\bf k},$\omega$) for the dispersing excitation that defines the FS
is much different for the LL than for a Fermi liquid
\cite{voit,schonhammer}. Since A({\bf k},$\omega$) can be measured by angle
resolved photoemission spectroscopy (ARPES), there has been strong motivation
for such studies of quasi-one-dimensional (q-1D) metals.  An unfortunate
complication for this line of research is that many q-1D metals display charge
density wave (CDW) formation and that strong CDW fluctuations involving
electron-phonon interactions above the CDW transition temperature can also
cause A({\bf k},$\omega$) to have NFL behavior which in some ways resembles
that of the LL \cite{LRA,mckenzie}.  For example, both scenarios predict a
substantial suppression of {\bf k}-integrated spectral weight near \ef,
bringing ambiguity to the interpretation of pioneering angle integrated
photoemission measurements \cite{baer-prl,baer-euro} which observed such a
weight suppression, and to subsequent ARPES studies \cite{ourbronzes,gweon} of
dispersing lineshapes in q-1D CDW materials.

Thus far ARPES studies of non-CDW q-1D metals have not obtained dispersing
lineshape data which could be compared meaningfully with many-body theories.
Most of the non-CDW q-1D metals are organic and for these metals {\bf
k}-integrated weight suppression near \ef\ occurs \cite{baer-euro}, but
dispersing features have not been observed \cite{grioniBS}. \lip\ is a 3D
material with bonding such that only q-1D bands define its FS. It is unusual as
a q-1D inorganic metal which appears to be free of strong electron-phonon
effects, as discussed further below, and which shows suppressed \ef\
photoemission weight.  An initial ARPES study \cite{smith} at 300K did not
resolve individual valence band features but did observe for a single broad
peak a general angle dependent shift and diminution of spectral weight which
enabled a q-1D FS to be deduced.  A second study \cite{gweon} obtained similar
data. A third study \cite{grioni} resolved valence band structure but the peak
dispersing to \ef\ was too weak in the spectra for its lineshape to be
discerned.

%%BEGIN_IN(tc_undo)
%%%BEGIN_IN(tc)
%%END_IN(tc_undo)
%%BEGIN_OUT(tc_undo)
\begin{figure}
%%END_OUT(tc_undo)
%%BEGIN_OUT(tc_undo)

%%END_OUT(tc_undo)
%%BEGIN_OUT(tc_undo)
\epsffile{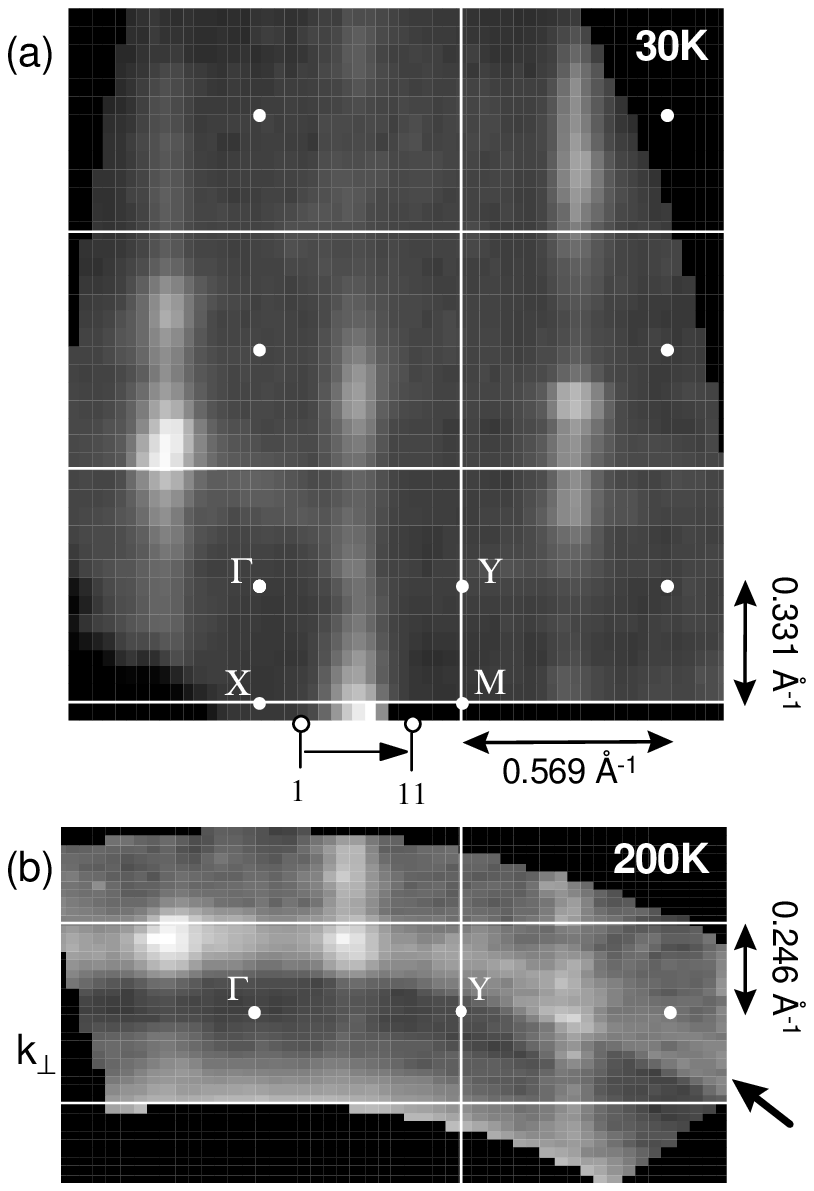}
%%END_OUT(tc_undo)
%%BEGIN_OUT(tc_undo)

%%END_OUT(tc_undo)
%%BEGIN_OUT(tc_undo)
\vspace{0.5cm}
%%END_OUT(tc_undo)
%%BEGIN_OUT(tc_undo)

%%END_OUT(tc_undo)
%%BEGIN_IN(tc_undo)
%%%%FIG1CAP
%%END_IN(tc_undo)
%%BEGIN_OUT(tc_undo)
\caption {Near-\ef\ intensity map of \lip. (a) \kpar\ plane projection for
h$\nu$=24 eV with variation of two detector angles.  (b) \kperp/$\Gamma$-Y
plane projection for varying h$\nu$=15-32 eV and one detector angle. The thick
arrow in (b) indicates the arc corresponding to h$\nu$=24 eV used in (a). In
both maps, image contrast has been enhanced by dividing the data by the data
heavily smoothed to retain only slowly varying cross-sectional dependences.}
%%END_OUT(tc_undo)
%%BEGIN_OUT(tc_undo)

%%END_OUT(tc_undo)
%%BEGIN_OUT(tc_undo)
\end{figure}
%%END_OUT(tc_undo)
%%BEGIN_IN(tc_undo)
%%%END_IN(tc)
%%END_IN(tc_undo)

Here we report the detailed non-Fermi liquid (NFL) lineshape of the dispersing
excitation which defines the FS for \lip.  Obtaining the lineshape data was
enabled by taking precautions to minimize photon-induced sample damage
\cite{damage} and by studying a region in {\bf k}-space where the near-\ef\
ARPES intensity is especially large, as determined by first making a {\bf
k}-space map of the ARPES intensity near \ef.  The properties of \lip\ strongly
argue that the NFL behavior has a purely electronic origin, giving this set of
data a special current importance. \lip\ displays metallic $T$-linear
resistivity $\rho$ and temperature independent magnetic susceptibility $\chi$
for temperatures down to $T_X \approx$ 24K, where a phase transition of unknown
origin is signaled by a very weak anomaly in the specific heat
\cite{schlenker}. As $T$ decreases below $T_X$, $\rho$ increases, but $\chi$ is
unchanged \cite{schlenker,chi}.  Most significant, infrared optical studies
which routinely detect CDW or spin density wave (SDW) gaps \cite{commonir} in
other materials, do not show any gap opening \cite{diGiorgi} for energies down
to 1 meV, setting an upper limit of (11.6/3.52)$\approx$3K for a mean field CDW
or SDW transition temperature.  Below $T_c \approx$ 1.8K the material is a
superconductor \cite{greenblatt}.  The properties of the 24K transition are not
consistent with CDW (or SDW) gap formation, and in any case, the small value of
$T_X$ permits the NFL ARPES lineshape to be studied from $T_X$ to nearly
10$T_X$, a temperature high enough that any putative q-1D CDW fluctuations
should be absent.  Comparing the data to the theoretical Luttinger liquid
lineshape, we identify significant similarities, but also find important
differences.

Single-crystal samples were grown by the electrolytic reduction technique
\cite{schlenker}.  The ARPES was performed at the Ames/Montana beamline of the
Synchrotron Radiation Center at the University of Wisconsin.  Samples oriented
by Laue diffraction were mounted on the tip of a helium refrigerator and
cleaved {\em in situ} at a temperature of 30K just before measurement in a
vacuum of $\approx 4\times 10^{-11}$ torr, exposing a clean surface containing
the crystallographic c- and q-1D b-axes.  Monochromatized photons of h$\nu$=24
eV were used to obtain the spectra reported here.  All the data are normalized
to the photon flux.  The instrumental resolution $\Delta$E and \ef\ were
calibrated with a reference spectrum taken on a freshly sputtered Pt foil.
$\Delta$E was 150 meV for the \ef\ intensity map and 50 meV for the energy
distribution curves (EDC's).  The angular resolution for the spectrometer was
$\pm 1^{\mathrm o}$, which amounts to $\pm 7$\% of the distance from $\Gamma$
to Y in the Brillouin zone.  The {\bf k}-space near-\ef\ intensity map was made
by detecting electrons over the range $\Delta$E=150meV, centered 50 meV below
\ef, and sweeping analyzer angles along two orthogonal directions relative to
the sample normal, in steps of 1$^{\mathrm o}$ for one angle and 2$^{\mathrm
o}$ for the other.  One can show \cite{ARPES} that such sweeps move the {\bf
k}-vector on a spherical surface with a radius which depends on the kinetic
energy and hence on the photon energy.  In an idealized geometrical
description, one observes the intersection of this spherical surface and the
FS.  In a more realistic spectroscopic description, the FS pattern is generated
because the intensity at \ef\ reaches a local maximum when a dispersing peak
passes sufficiently near the angle/energy resolution window, for given
temperature and peak lineshape.  Because of translational invariance parallel
to the sample surface the photohole momentum parallel to the surface \kpar\ is
the same as that of the photoelectron and so is determined unambiguously by the
analyzer angles and the kinetic energy of the photoelectron \cite{ARPES}.  To
deduce the perpendicular photohole momentum (\kperp) the surface potential
change must be modeled and in making our maps we have used a standard ansatz
\cite{ARPES} of free photoelectron bands offset by an inner potential to which
we give a nominal value of 10 eV.

Fig.\ 1(a) shows the projection onto the \kpar\ plane of our near-\ef\
intensity map made at a temperature of 30K by varying both analyzer angles for
fixed h$\nu$=24 eV.  $\Gamma$-Y and $\Gamma$-X are the b* and c* directions,
respectively.  Fig.\ 1(b) shows the projection onto the \kperp/$\Gamma$-Y plane
of a map made at 200K by fixing one analyzer angle, while varying the other
angle and also the photon energy.  The spherical arcs for each photon energy
are easily seen, and an arrow shows the arc corresponding to the fixed photon
energy of the map of Fig.\ 1(a).  The straightness of the FS segments in both
maps shows that this material fulfills very well the band theory prediction of
being q-1D \cite{band,kfdetail}.  The Fermi wave-vector k$_F$ defined by the
center of the left hand FS segment is 2k$_F$$\approx$$0.57 $\AA$^{-1}$,
somewhat larger than the band theory \cite{band,kfdetail} value of $0.51$
\AA$^{-1}$.  Most significant for the rest of the paper is the existence of
bright spots where the ARPES matrix element for the states near \ef\ is maximum
and where the dispersing peak lineshape can best be studied.

Fig.\ 2(a) shows a sequence of spectra taken at 200K $\approx$ 8$T_X$ along a
line $0.06 $\AA$^{-1}$ below an X-M Brillouin zone boundary and passing through
the FS at one of the bright points, as indicated in Fig.\ 1(a). Over the
corresponding k-range along $\Gamma$-Y, the calculation of Fig.\ 2(b) shows two
bands merging and crossing \ef\ together. We identify the two dispersing peaks
of Fig.\ 2(a) with these two bands, since both the calculation and our q-1D FS
image show that the two bands disperse very weakly along $\Gamma$-X. The
calculated bands which do not cross \ef\ are very weak for the special path of
Fig.\ 2(a), but can still be seen as a small peak or general humping $\sim 400$
meV below \ef\ in spectra 4 to 11. These bands are easily seen in other
spectra, e.g.\ along $\Gamma$-X and $\Gamma$-Y. Thus we find a good general
agreement with band theory except that the bandwidth is about twice the
calculated value, as has been found for other molybdenum bronzes
\cite{gweon,gweon2}. Since LL models assume linear dispersion around \ef, it is
noteworthy that this aspect of the band theory is observed over an energy range
of 200 meV for one band and 500 meV for the other. Of greatest interest is the
detailed lineshape of the dispersing peak defining the FS. It moves toward \ef\
from about 250 meV away until the leading edge shows a point of closest
approach, after which the intensity then drops.  Within the experimental
resolution, very little intensity develops at \ef.  Fig.\ 3(a) shows the
spectra overplotted so as to emphasize the defining behavior of the leading
edge of the lineshape.  In spectra 2 through 5 one sees the leading edge shift
toward \ef\ up to a certain limit, ``the wall,'' and in spectra 5 through 10
one sees the intensity fall, first without a change in the leading edge, and
then accompanied by a shift of the leading edge away from \ef. A set of spectra
taken at 50K is identical with respect to all these features.

%%BEGIN_IN(tc_undo)
%%%BEGIN_IN(tc)
%%END_IN(tc_undo)
%%BEGIN_OUT(tc_undo)
\begin{figure}
%%END_OUT(tc_undo)
%%BEGIN_OUT(tc_undo)

%%END_OUT(tc_undo)
%%BEGIN_OUT(tc_undo)
\epsffile{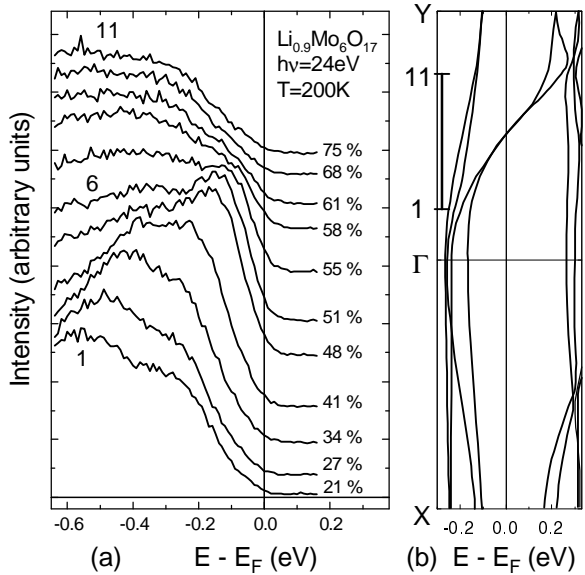}
%%END_OUT(tc_undo)
%%BEGIN_OUT(tc_undo)

%%END_OUT(tc_undo)
%%BEGIN_OUT(tc_undo)
\vspace{0.5cm}
%%END_OUT(tc_undo)
%%BEGIN_OUT(tc_undo)

%%END_OUT(tc_undo)
%%BEGIN_IN(tc_undo)
%%%%FIG2CAP
%%END_IN(tc_undo)
%%BEGIN_OUT(tc_undo)
\caption {(a) ARPES spectra showing FS crossing along the path, marked with an
arrow in Fig.\ 1(a), that passes through the bright spot of the image. For each
spectrum, the corresponding momentum value parallel to $\Gamma$Y is given in
percentage of the length of $\Gamma$Y. (b) Tight binding band calculation
[\ref{band}] showing bands along X--$\Gamma$--Y. The bar shows the range of
k$_{\Gamma Y}$ explored in (a).}
%%END_OUT(tc_undo)
%%BEGIN_OUT(tc_undo)

%%END_OUT(tc_undo)
%%BEGIN_OUT(tc_undo)
\end{figure}
%%END_OUT(tc_undo)
%%BEGIN_IN(tc_undo)
%%%END_IN(tc)
%%END_IN(tc_undo)

In the absence of any LL lineshape theory including interactions between two
bands, we apply lineshapes calculated for the one-band Tomonaga-Luttinger (TL)
model to the two degenerate bands crossing \ef . Fig.\ 3(b) shows TL lineshapes
for a spin independent repulsive interaction \cite{schonhammer} and singularity
index $\alpha=0.9$.  The thick lines are spectra including our angle and energy
resolutions.  The thin lines accompanying two of the spectra show the purely
theoretical curves without including the experimental resolutions.  The
k-values and format are exactly the same as for Fig.\ 3(a).  Before discussing
the considerable similarity to the experimental data for the behavior of the
leading edge, we first describe the generic theoretical features. The LL has no
single particle excitations, and the removal or addition of an electron results
entirely in the generation of combinations of collective excitations of the
spin and charge densities, known as spinons and holons, respectively.  In this
TL model the spinon dispersion is that of the underlying band, v$_F$k with
Fermi velocity v$_F$, and the holon dispersion is $\beta$v$_F$k where $\beta$
depends on $\alpha$ and is $>$ 1.  For the lower group of spectra, with k
inside the FS, there is an edge singularity onset at a non-zero low energy and
then a rise to a power law singularity peak at higher energies.  These sharp
features are greatly broadened by the experimental resolutions and, except for
the slight shoulder of curve 2, the spinon features of the theory curves are
simply the leading edges of the lineshapes.  The movements with k of the low
energy onset and of the peak reflect the dispersions of the spinons and holons,
respectively.  That the onset occurs at a non-zero energy for k$\neq$k$_F$ is a
direct consequence of the restrictive kinematics of 1-D.  For the four lowest
members of the upper set of curves, k lies outside the FS.  The k-dependence of
the non-zero singular energy onset in this case reflects the holon dispersion.

%%BEGIN_IN(tc_undo)
%%%BEGIN_IN(tc)
%%END_IN(tc_undo)
%%BEGIN_OUT(tc_undo)
\begin{figure}
%%END_OUT(tc_undo)
%%BEGIN_OUT(tc_undo)

%%END_OUT(tc_undo)
%%BEGIN_OUT(tc_undo)
\epsffile{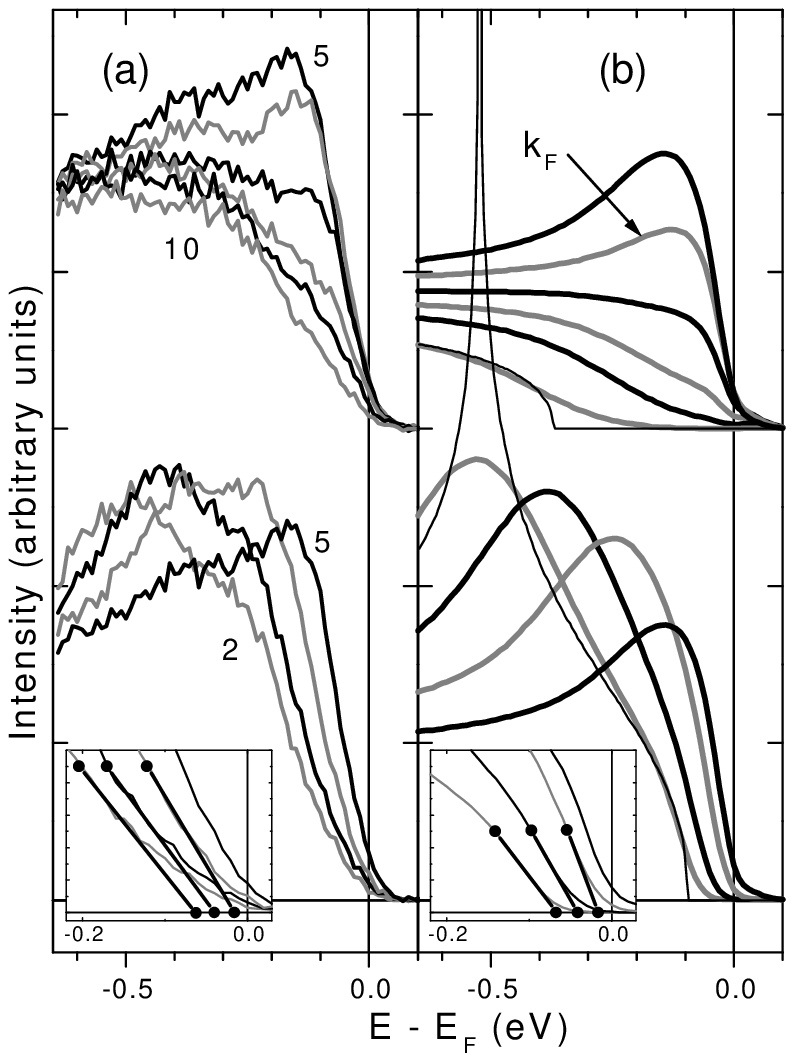}
%%END_OUT(tc_undo)
%%BEGIN_OUT(tc_undo)

%%END_OUT(tc_undo)
%%BEGIN_OUT(tc_undo)
\vspace{0.5cm}
%%END_OUT(tc_undo)
%%BEGIN_OUT(tc_undo)

%%END_OUT(tc_undo)
%%BEGIN_IN(tc_undo)
%%%%FIG3CAP
%%END_IN(tc_undo)
%%BEGIN_OUT(tc_undo)
\caption {(a) Spectra of Fig.\ 2(a) replotted to emphasize the ``wall''
behavior, described in text.  Inset shows a detailed view of the spectral line
shape approaching \ef, with lines drawn to emphasize a hint of 1D onset
behavior. (b) Tomonaga-Luttinger (TL) model spectra, calculated to be compared
with (a), as described in text. Inset shows the spinon edge singularity
onsets.}
%%END_OUT(tc_undo)
%%BEGIN_OUT(tc_undo)

%%END_OUT(tc_undo)
%%BEGIN_OUT(tc_undo)
\end{figure}
%%END_OUT(tc_undo)
%%BEGIN_IN(tc_undo)
%%%END_IN(tc)
%%END_IN(tc_undo)

We now discuss the choice of parameters and the comparison to experiment, for
which we associate the spectral peaks with the rapidly dispersing holon
features and the leading edges with the slowly dispersing spinon features. We
consider a range of $\alpha > 1/2$ because for $\alpha < 1/2$ the low energy
edge singularity takes the form of a peak which is obviously not present in the
data.  Each $\alpha$ determines a $\beta$ and v$_F$ is chosen so that
$\beta$v$_F$k matches the experimental peak movements, linear to $\approx 500$
meV below \ef\ for one peak, but only $\approx 200$ meV below \ef\ for the
other, so that the lowest energy peak in data curves 1 to 3 has no theoretical
counterpart. One finds that for the broadened spectra, as $\alpha$ increases
from 1/2, (a) the peak maximum as k approaches k$_F$ decreases more rapidly,
and (b) the amount of \ef\ weight relative to the spectrum maximum in the
k=k$_F$ spectrum decreases.  As expected in the TL theory
\cite{voit,schonhammer}, we have observed a power law onset at \ef\ in a
measurement of the angle-integrated photoemission spectrum, from which we
deduce $\alpha \approx$ 0.6, nicely greater than 1/2.  For $\alpha$ = 0.6
($\beta= 4$) the behavior of (a) is similar to experiment but the value for (b)
is about twice the experiment value of $\approx$ 16\%.  For $\alpha$ = 0.9
($\beta= 5$), it is noticeable that the behavior of (a) is faster than in
experiment, but the fractional amount of \ef\ weight for the k=k$_F$ spectrum
is only slightly greater than in experiment.  With the choice $\alpha$ = 0.9
and $\hbar$v$_F$ = 0.7 eV\AA\ \cite{rc}, the theory curves reproduce
semiquantitatively the variation of the leading edge in spectra 2 to 5, the
``wall'' behavior in spectra 5 to 7, the loss of a peaky upturn at \ef\ from
spectrum 6 to 7 as k passes beyond k$_F$, and qualitatively the movement of the
leading edge away from \ef\ for spectra 8 to 10.  The agreement of the
intercepts given by the straight line extrapolations shown in the two insets
indicates a remnant in the data of the theoretical onset behavior of 1-D
kinematics, and even a semi-quantitative agreement with the $\beta$ value.  The
general goodness of the agreement for spectra 5 through 7 leads us to take the
value of 2k$_F$=0.59 \AA$^{-1}$ from spectrum 6 as a better determination of
2k$_F$ than the slightly smaller value deduced above from the center of the FS
image.

Looking in more detail, differences can be seen.  First, considering the insets
of Fig.\ 3, the amount of experimental weight in the energy range from \ef\ to
the theory onset definitely exceeds that for the corresponding broadened theory
curve.  This could reflect the ultimate 3-D character of the material relaxing
the restrictive 1-D kinematics, consistent with the increasing magnitude of the
disagreement as {\bf k} moves further from the FS and the available phase space
increases.  We also report that the only difference between the spectra at 200K
and 50K is a subtle change at lower temperature such that the leading edge
extrapolates more to \ef. At present this small temperature dependence is a
tentative finding which requires further study, but might hint at a departure
from LL behavior, perhaps an increased 3-D character, due to some lingering
effects of whatever processes are important in the phase transition at $T_X$.
In any case, we note that this temperature dependence is opposite to that
expected \cite{mckenzie} for the case of a pseudogap associated with gap
formation (e.g.\ CDW or SDW) at $T_X$.  Second, the magnitude of the edge
movement for experimental spectra 8 to 10 is much less than in the theory,
probably due to the interfering presence in the spectra of the contributions
from the two bands further below \ef.  Thus the detailed differences can
plausibly be attributed to the oversimplifications of the TL model, e.g.\ its
one-band nature and its strict 1-D character, relative to the experimental
situation. The fact that our $\alpha$ value is much larger than the value 1/8
for the 1-D Hubbard model could also be a consequence of some 3-D coupling
\cite{theory3d}.
  
In summary we have presented spectra which are currently unique in showing the
lineshape of the dispersing excitation that defines the FS for interacting
electrons in a q-1D non-CDW metal.  We have compared the data to the lineshape
in the TL model of the LL.  Although there are important differences in detail,
nonetheless there is a remarkable similarity between theory and experiment for
the anomalous behavior of the leading edge of the lineshape.  In the TL model
this behavior has its origin in the underlying charge-spin separation of the LL
scenario. Previous ARPES reports \cite{kim} of charge-spin separation have been
for q-1D materials where a Mott-Hubbard insulator precludes the LL. This is the
first such report for a q-1D metal and provides strong motivation for further
study of \lip\ using other techniques.

Work at U-M was supported by the U.S.\ Department of Energy (DoE) under
contract No.\ DE-FG02-90ER45416 and by the U.S. National Science Foundation
(NSF) grant No.\ DMR-94-23741. Work at the Ames lab was supported by the DoE
under contract No.\ W-7405-ENG-82. The Synchrotron Radiation Center is
supported by the NSF under grant DMR-95-31009.


\begin{thebibliography}{99}
\bibitem{kfdetail} Ref.\ \ref{smith} found larger dispersion perpendicular
to the sample surface than we do and a different 2k$_F$ $\approx$ 0.7
\AA$^{-1}$. Both differences can be attributed to our better resolution of the
near-\ef\ peak which defines the FS.
\bibitem{rc} For the large k,$\omega$ range of the linear dispersion, the universal
spectral function used here is valid if the potential range parameter $r_c <
0.7$\AA \cite{schonhammer}.  The $r_c$ dependence is then too weak to
discern in the comparison to the data and we conservatively chose $r_c =
0.1$\AA.
\end{thebibliography}
\end{document}